\documentclass[%
 reprint,
 showkeys,
 amsmath,amssymb,
 aps,
]{revtex4-2}
\usepackage{float}
\usepackage{booktabs}
\usepackage{graphicx}
\usepackage{dcolumn}
\usepackage{bm}
\usepackage[table,xcdraw,dvipsnames]{xcolor}
\usepackage{multirow}

\usepackage{amsmath}
\usepackage{xr-hyper} 
\usepackage{colortbl}
\usepackage{placeins}
\usepackage[font=small,skip=4pt]{caption}
\usepackage{hyperref}
\hypersetup{
colorlinks=true,
linkcolor=blue,
citecolor=blue,
urlcolor=blue
}

\begin{document}

\title{
Development and application of a machine learning potential for diamond surfaces and their sliding interfaces with silicon nitride and silica: effect of passivation, orientation and load
}

\author{Enrico Pedretti$^{1}$}
\author{Maria Clelia Righi$^{1}$}\email{clelia.righi@unibo.it}

\affiliation{
$^{1}$Physics and Astronomy Department, University of Bologna,\\
 Viale Berti Pichat 6/2, 40127, Bologna (Italy)\\
}

\begin{abstract}

Friction and wear of diamond are controlled by its surface chemistry, which determines whether covalent bonds form across the sliding interface. First-principles simulations have explored the role of surface orientation and passivation, but are limited to a few hundred atoms and tens of picoseconds. To bridge this gap, we develop a reactive machine learning interatomic potential for the C/H/N/O/Si chemical space, covering diamond, silicon nitride, silicon oxide, their surfaces, interfaces, and environmental molecules. The potential is trained on a large dataset of diverse structures and is tested on multiple physical properties, including \textit{ab initio} reaction barriers on diamond surfaces, and structural/mechanical properties of amorphous Si$_3$N$_4$ and SiO$_2$. We then simulate  sliding of diamond--Si$_3$N$_4$ interfaces with four surface orientations, clean and H/OH-passivated terminations, and normal loads from 0.1 to 23 GPa. We identify two different regimes: one where sliding is localized at the interface, and shear stress is proportional to the interfacial bond density, and another where shear is distributed inside the softer material, and adhesion due to full interfacial bonding dominates. These results show how passivation, surface orientation and load together determine interfacial bonding and friction, with implications for the environmental sensitivity of diamond wear. More broadly, the potential enables reactive simulations of diamond, silicon nitride and silica with near-\textit{ab initio} accuracy at length and time scales beyond first-principles methods, providing a useful tool for future studies of wear at plowing asperities, chemical-mechanical polishing, and properties of amorphous ceramic films.

\end{abstract}


\maketitle
\thispagestyle{titlepage}
\section{Introduction}

Diamond is widely used as protective coating and functional material in applications ranging from cutting tools and bearings to micro-electro-mechanical systems~\cite{DIAM-liu2010preventing,DIAM-wheeler2018applications,DIAM-araujo2021diamond,DIAM-hainsworth2007diamond,DIAM-auciello2021review,DIAM-erdemir2006tribology}. Particularly stringent requirements arise in the semiconductor industry, where diamond-coated components are used for the precision positioning of wafers, and friction and even sub-nanometer wear affect performance. In all these applications, the tribological behavior of diamond depends strongly on the environment: in humid air it can slide with very low friction and wear, whereas in dry or vacuum conditions both increase markedly~\cite{DIAM-konicek2008origin,DIAM-konicek2012influence}.

This sensitivity originates from the chemistry of the diamond surface. In dry or vacuum conditions, unsaturated surface carbon atoms form covalent bonds with the counter-body, which increase adhesion and friction~\cite{DIAM-milne2019covalent,DIAM-cutini2023adhesion} and constitute the first step of atom-by-atom wear~\cite{DIAM-jacobs2013nanoscale,DIAM-gotsmann2008atomistic,DIAM-lin2021high}. In humid air, instead, the dissociation of O$_2$ and H$_2$O can passivate these sites with O, H and OH groups, which prevent interfacial bonding~\cite{DIAM-tran2022}. The specific behaviour also depends on the exposed surface, since the reactivity of diamond toward both the counter-body and environmental molecules varies strongly with orientation and reconstruction~\cite{DIAM-tran2022}. 

These aspects have been extensively explored through first-principles simulations. Static calculations, in which a surface atom is pulled away along a predefined path, demonstrated that carbon is most easily removed from C(110) and least easily from the Pandey-reconstructed C(111)~\cite{Pandey1981new,Vanderbilt1984total,DIAM-ta2023effects,giaremis2025exploring}. Moreover, the reduction of adhesion and friction by surface passivation has been explored by \textit{ab initio} molecular dynamics (AIMD) simulations of diamond sliding against silica, where the absence of passivation lead to the formation of bonds to the counter-surface through Si--O--C and Si--C bridges~\cite{DIAM-cutini2023adhesion,DIAM-Ta2023,giaremis2025tribological}. Combining AIMD with precision wear experiments, we also recently showed that the same mechanism controls the wear of micro-crystalline diamond (MCD) coatings sliding against amorphous Si$_3$N$_4$~\cite{leriche2025passivation}.

These approaches, however, have clear limitations: as AIMD covers a few tens of picoseconds of sliding in systems of a few hundred atoms, it can capture the onset of interfacial bonding, but neither the steady state reached after run-in nor the plastic response of the counter-body. Therefore, the role of surface orientation, reconstruction, passivation and load in determining interfacial bonding and friction over the range of pressures found in a multi-asperity contact is still not fully understood, and addressing it requires \textit{ab initio} accuracy over much longer time scales and system sizes.

Machine learning interatomic potentials (MLPs) can bridge this gap, providing near-DFT accuracy at a small fraction of the computational cost~\cite{ZDDP-batatia2022mace,ZDDP-batatia2025design}. For the present tribological problem, however, no existing potential covers the required chemistry. Accurate MLPs exist for carbon in all its phases~\cite{deringer2017machine,rowe2020accurate}, and a DeepMD-based potential has recently been proposed for hydrogenated diamond and amorphous carbon under deposition, friction and fracture~\cite{Chen2025transferable}, but these include neither oxygen nor interfaces with silicon nitride or silica. Conversely, dedicated potentials exist for silicon nitride~\cite{xu2023high,nayak2025accurate,Milardovich2023machine}, including one for Si$_3$N$_4$ tribology containing O and H but tuned for water and ethylene glycol lubrication~\cite{kudo2024neural}, and for silica--water systems~\cite{Roy2024learning}. None of these can describe a loaded, sheared diamond--ceramic interface in which passivating species (H, OH, O) take part in the interfacial chemistry. Universal ``foundation'' models~\cite{ZDDP-batatia2023foundation,ZDDP-matpes} span the full C/H/N/O/Si element set, but require fine-tuning to reach the accuracy needed for surface energetics and reaction barriers under tribological conditions.

In this work, we develop a general-purpose machine learning potential for the C/H/N/O/Si chemical space, designed to describe diamond, diamond-like carbon (DLC), silicon nitride, silica and silicon in crystalline and amorphous form, their surfaces, environmental molecules, and their mutual interfaces under load and shear. Two MACE~\cite{ZDDP-batatia2022mace,ZDDP-batatia2025design} models are trained on the same DFT dataset: a highly accurate model fine-tuned from the MACE-MATPES foundation model~\cite{ZDDP-batatia2023foundation,ZDDP-matpes} (MACE-ft), and a lighter model (MACE-small) that trades a small amount of accuracy for a large gain in speed and GPU memory efficiency. Beyond energies and forces, both models are validated on reaction barriers and on structural and mechanical properties, and their applicability beyond tribology is demonstrated through tensile tests of amorphous Si$_3$N$_4$ and SiO$_2$ and the CVD-like growth of a silicon nitride film~\cite{zheng2013deposition,Lacava2017sirich}.

Finally, we employ the potential to study diamond--Si$_3$N$_4$ interfaces sliding under load, systematically covering four diamond orientations (C(001), C(110), C(111), and Pandey-reconstructed C(111)), clean and H/OH-passivated terminations, and normal loads from 0.1 to 23 GPa. These simulations allow us to establish a quantitative relation between load, interfacial bond formation and frictional shear stress, and to characterize, through the analysis of velocity profiles, a transition from interfacial slip to distributed plastic shear within the amorphous nitride once chemical bonding saturates the interface. The results extend the \textit{ab initio} picture of passivation-controlled interfacial bonding~\cite{leriche2025passivation} to steady-state sliding over the full range of contact pressures and, since interfacial bonding is also the first step of atom-by-atom wear, are relevant to the interpretation of diamond wear experiments.

\section{Methods}

\subsection{MLP training}
This work employs the MACE machine learning potential architecture~\cite{ZDDP-batatia2022mace,ZDDP-batatia2025design}, selected for its excellent accuracy in describing complex and reactive systems. As for any MLP, the most crucial ingredient is the DFT dataset of energies and forces used for training, which should contain a large number of diverse and independent atomic structures, so that the configuration space explored during the MD simulations is sufficiently and homogeneously sampled.
The dataset was constructed through the Strategic Configuration Sampling (SCS) active-learning framework developed in our group~\cite{ZDDP-pacini2025accelerating}, which employs an iterative process to produce a complete DFT dataset over a wide range of conditions, and allows to automatically build structures of bulks, slabs, and interfaces.

\begin{figure*}[!htb]
\centering
\includegraphics[width=\textwidth]{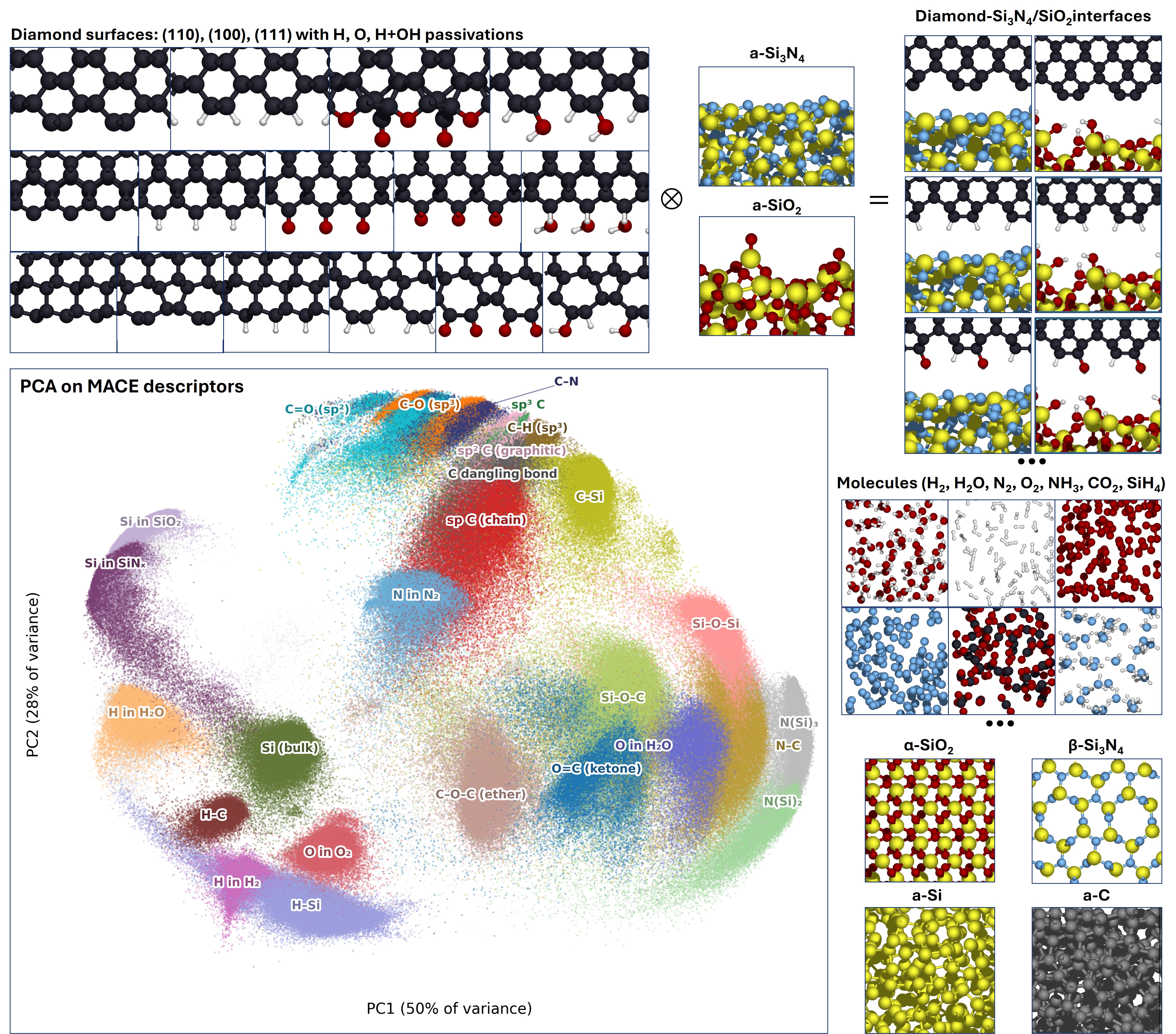}
\caption{Overview of the dataset. The most relevant systems are shown, together with the PCA dimensionality reduction on learned MACE descriptors across the dataset. The clustering resulting from the PCA projection has a clear correspondence with (independent) coloring and labeling based on atomic coordination, confirming the ability of the potential to correctly distinguish the different local environments.}
\label{fig:dataset}
\end{figure*}

The dataset was designed following a ``building blocks + interactions'' strategy: we first included the individual materials and molecules in a wide range of thermodynamic conditions, and then their combined systems under conditions representative of the target tribological applications. In particular, the dataset comprises:
\begin{itemize}
    \item \textbf{Carbon phases}: bulk diamond and graphite; bulk DLC with densities in the range 1.5--3.5 g/cm$^3$ and hydrogen content from 0 to 30\%at., generated by melt-quench simulations up to 7000 K; DLC slabs annealed up to 1000 K; graphitization processes.
    \item \textbf{Diamond surfaces}: C(001), C(110), C(111) and Pandey-reconstructed C(111)~\cite{Pandey1981new,Vanderbilt1984total}, each clean and passivated with H, H+OH, and O, at both 50\% and 100\% coverage. Oxygen passivations were constructed with different geometries, with O in ketone and ether configurations, as in a previous work of our group~\cite{DIAM-ta2023effects}, and including the ketone--ether coexistence for C(110) reported in Ref.~\cite{Chaudhuri2022coexistence}. For C(001), the 50\% O coverage features ether bridges on top of the reconstructed dimers, while at 100\% coverage the reconstruction is lifted, with either ether bridges or ketone terminations; the Pandey-reconstructed C(111) hosts ethers at 50\% coverage and ketones/peroxides at 100\%. All H+OH passivations were obtained by imagining the dissociation of water molecules, i.e. with equal amounts of H and OH groups paired close together. Stepped surfaces, with steps along two perpendicular directions, were also included. All surfaces were sampled by NVT temperature ramps up to 1000 K.
    \item \textbf{Molecules}: H$_2$, O$_2$, H$_2$O, CO, CO$_2$, N$_2$, NH$_3$, NO, NO$_2$, OH and SiH$_4$ in periodic boxes with temperature ramps up to 5000 K, together with N$_2$+H$_2$ and N$_2$+O$_2$+H$_2$O mixtures, and the reaction of NH$_3$+SiH$_4$ mixtures at high temperature to form silicon nitride clusters and H$_2$.
    \item \textbf{Silicon nitride}: bulk melt-quenches up to 5000 K and slab anneals up to 2000 K, spanning a wide range of stoichiometries and hydrogen contents, SiN$_x$H$_y$ with $x = 0.1$--$1.5$ (from almost pure Si to N excess) and $y = 0$--$0.3$, with the cell volume set to match the experimental density, interpolated in the range 2.3--3.2 g/cm$^3$ for $x=0$--$1.5$; uniaxial compressions up to 23 GPa.
    \item \textbf{Silica and silicon}: crystalline SiO$_2$ under heating; amorphous silica (bulk and slabs) with densities from 2.0 to 2.4 g/cm$^3$ and hydrogen contents of 0 and 10\%at., generated by melt-quenches up to 5000 K; crystalline Si bulk under heating; amorphous Si from melt-quench; and the Si(100) surface under heating and oxidation by O$_2$.
    \item \textbf{Surface--molecule interactions}: the clean C(001), C(110), C(111) and Pandey-reconstructed C(111) diamond surfaces, plus amorphous silica (clean and silanol-terminated), Si$_3$N$_4$ and DLC surfaces, interacting with O$_2$, H$_2$, H$_2$O and N$_2$ molecules up to 1500 K (including NPT compression runs), in order to spontaneously sample dissociation events and passivation processes.
    \item \textbf{Interfaces}: diamond--Si$_3$N$_4$ and diamond--silica interfaces, for all the aforementioned diamond surfaces (orientations, passivations, and steps), under normal loads of 0.1, 1, 10 and 23 GPa, with sliding along the two perpendicular in-plane crystallographic directions at velocities up to 25 m/s and temperatures up to 600 K, and pull-off (interface separation) trajectories; the same interfaces with intercalated H$_2$, O$_2$, H$_2$O and N$_2$ molecules and their mixtures; plus DLC--Si$_3$N$_4$/DLC--silica interfaces.
\end{itemize}
The final dataset comprises 253 systems for a total of 30423 configurations ($\sim$14.8M atomic environments), an overview of which is presented in Fig.~\ref{fig:dataset}.

To also confirm the ability of our MACE potential to correctly recognize the different atomic environments, we performed PCA dimensionality reduction on the learned descriptors, which were computed extracting only the invariant part of the first message passing layer (128 scalars for each descriptor). The first two components from the PCA are plotted in Fig.~\ref{fig:dataset} (showing only a randomly-sampled 10\% of the dataset). Independent classification of environments, based on local coordination, was used to label the different parts of the dataset, showing a clear correspondence between clusters in the PCA projection of MACE descriptors and the different local atomic environments.

The reference DFT energies and forces for the dataset were computed with the plane-wave code Quantum ESPRESSO~\cite{ZDDP-Giannozzi2009,ZDDP-Giannozzi2017,ZDDP-Giannozzi2020}, using the PBE exchange-correlation functional~\cite{ZDDP-Perdew1996} and PAW pseudopotentials~\cite{ZDDP-blochl1994projector}, with a kinetic energy cutoff of 60 Ry for the wave functions and 480 Ry for the charge density, and a Gaussian smearing of 0.01 Ry. Spin polarization was used to capture reactive events of bond breaking and formation, and dispersion interactions were included through the Grimme-D3 correction scheme~\cite{ZDDP-d3}.

Two neural network potentials were trained on this dataset:
\begin{itemize}
    \item \textbf{MACE-ft}: obtained by (multi-head) fine-tuning the pretrained universal potential MACE-MATPES-PBE-0~\cite{ZDDP-matpes} on our dataset. This is the heavier model, with 128 channels and cutoff radius of 6 Å (effective 12 Å due to message passing).
    \item \textbf{MACE-small}: a much lighter model trained from scratch on our dataset, with 64 channels and a cutoff radius of 4 Å (effective 8 Å due to message passing). Its accuracy is only slightly worse than that of MACE-ft (as quantified in the next section), making it an optimal compromise for large-scale simulations.
\end{itemize}

The computational advantage of the small model is substantial, mainly owing to its shorter cutoff radius, which greatly reduces the number of neighbors entering the message passing. This enables reaching up to 2.7 Msteps/day on A100(64 GB) GPUs with $\sim$3000 atoms/GPU, and 1 Msteps/day for a system with one million atoms distributed over 80 A100 GPUs ($\sim$13000 atoms/GPU), whereas models of the size of MACE-ft exceed the available GPU memory beyond roughly 5000-6000 atoms/GPU on the same hardware.

All the production simulations presented in this section were therefore performed with the MACE-small model, while MACE-ft serves as the accuracy reference and remains available for applications demanding the highest fidelity.

\subsection{Testing the MLPs' accuracy}

\begin{figure*}[!htb]
\centering
\includegraphics[width=\textwidth]{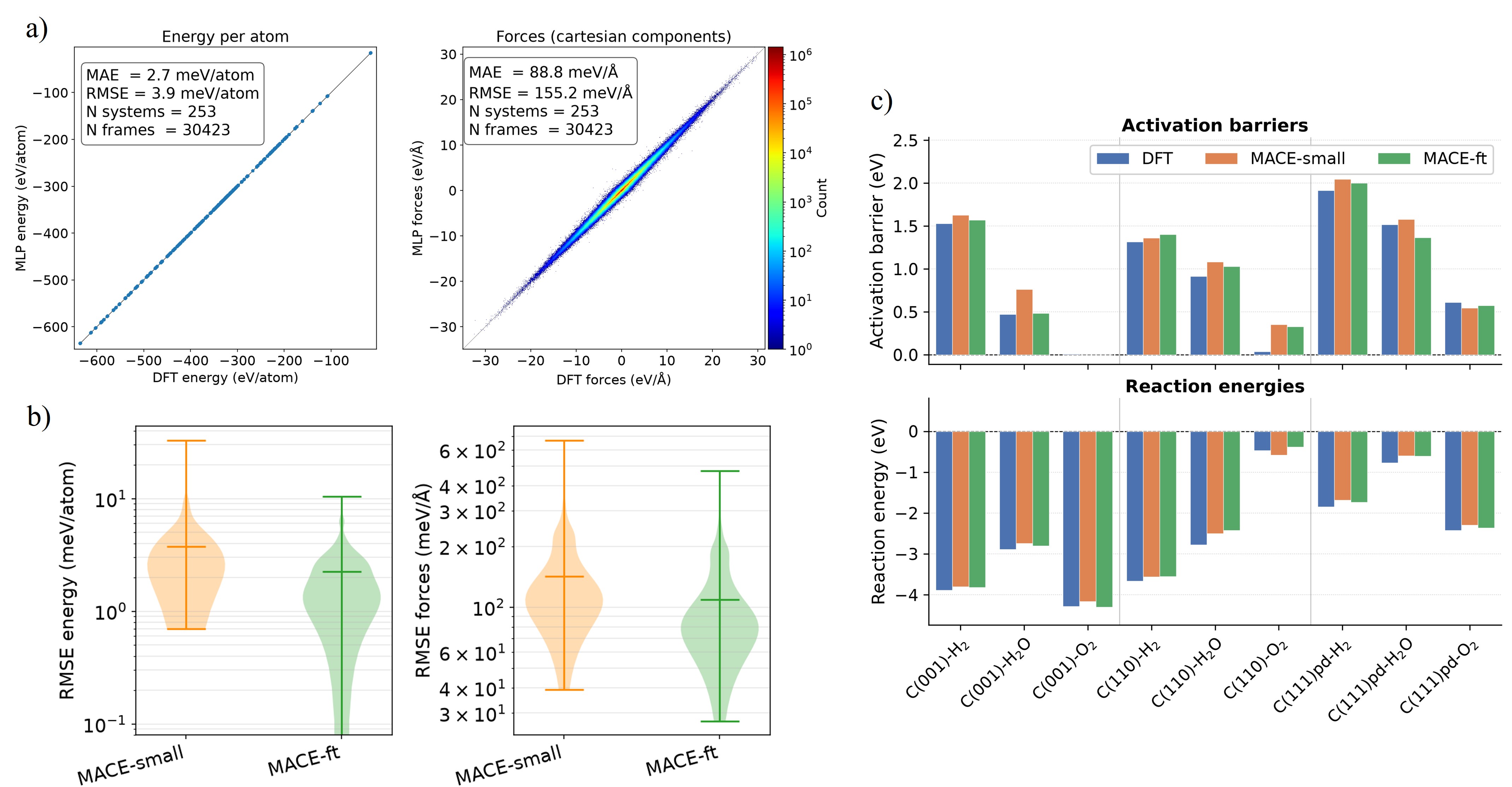}
\caption{Accuracy metrics for the ML potential(s): a) parity plots of energies and forces for the MACE-small model used for the simulations; b) violin plots with per-system distribution of energy and force RMSE, showing that except for some outliers at high energies and temperatures, many systems are below 1.5 meV/atom RMSE of energy and 70 meV/\AA\ RMSE of forces for the MACE-ft model, and 3 meV/atom RMSE of energy and 100 meV/\AA\ RMSE of forces for the MACE-small model; c) physical tests on activation barriers and reaction energies on selected dissociation reactions of hydrogen, oxygen and water on diamond surfaces (detailed paths in Fig.~\textbf{S3}).}
\label{fig:ml_accuracy}
\end{figure*}

To assess the accuracy of the trained models, we first compared the predicted energies and forces against the DFT reference across the entire dataset. The parity plots for the MACE-small model, reported in Fig.~\ref{fig:ml_accuracy}a, show excellent agreement, with a mean absolute error (MAE) of 2.7 meV/atom on energies (RMSE of 3.9 meV/atom) and 88.8 meV/Å on force components (RMSE of 155.2 meV/Å) over the 253 systems and 30423 frames of the dataset. Considering the extremely wide range of temperatures (up to 7000 K), pressures, densities and chemical environments contained in the dataset, these errors fall well within the accepted accuracy range for reliable interatomic potentials~\cite{ZDDP-liu2023discrepancies,ZDDP-zuo2020performance}. The per-system distributions of the errors, shown as violin plots in Fig.~\ref{fig:ml_accuracy}b, confirm that, apart from a few outlier systems associated with the highest temperatures and energies, most systems fall below 1.5 meV/atom and 70 meV/Å (RMSE of energies and forces) for MACE-ft, and below 3 meV/atom and 100 meV/Å for MACE-small. The complete system-by-system parity plots for MACE-small and the comparison of the per-system force RMSE between the two models are reported in the Supplementary Information (Figs.~\textbf{S1} and ~\textbf{S2}): MACE-ft is consistently the more accurate of the two, but the accuracy of MACE-small remains remarkably close, justifying its use for the production simulations.

Beyond errors on energies and forces, which quantify the agreement on the dataset itself, we performed a series of tests on physically meaningful quantities. First, we tested the dissociation reactions of H$_2$, H$_2$O and O$_2$ on the C(001), C(110) and Pandey-reconstructed C(111) diamond surfaces, considering several reaction pathways from our previous work \cite{DIAM-tran2022} computed with the nudged elastic band (NEB) method~\cite{ZDDP-henkelman2000improved}, here recomputed statically with the two MLPs. The resulting activation barriers and reaction energies for the considered pathways are compared in Fig.~\ref{fig:ml_accuracy}c, while the full reaction profiles are reported in the SI (Fig.~\textbf{S3}). Both models reproduce the DFT barriers and reaction energies with excellent accuracy in almost all cases, with deviations up to a few tenth of eV. It should be noted that the \textit{ab initio} values are not necessarily to be taken as the ``true'' unique reaction pathways: multiple pathways are possible, starting from either physisorbed or chemisorbed initial configurations, and these calculations are to be considered as benchmarks for the ML potentials rather than as a complete characterization of the surface reaction kinetics.

Second, as a test directly relevant for wear, we considered the peeling of a carbon atom chain from the C(110) surface by vertical pulling of a silicon nitride cluster initially bonded to the surface, i.e. the atomistic wear process characterized at the DFT level in our previous work~\cite{leriche2025passivation}. The energy profiles along the pulling coordinate, recomputed with the two MLPs on the DFT configurations, are reported in the SI (Fig.~\textbf{S4}): both models accurately capture the sequence of carbon detachment events and the associated energy variations, confirming that the potentials can be trusted for the description of the bond-breaking processes underlying atom-by-atom diamond wear.

Finally, we verified the description of the vibrational and elastic properties of the crystalline phases by computing the phonon band structures and densities of states of diamond, $\beta$-Si$_3$N$_4$ and $\alpha$-quartz SiO$_2$ with the MACE-small model (Fig.~\textbf{S5}), obtaining a good agreement with the reference MaterialsProject data~\cite{Horton2025,Jain-2013} across the whole spectrum.

\subsection{Molecular dynamics simulations}
All the MD simulations presented in this section were performed with the LAMMPS software~\cite{ZDDP-THOMPSON2022108171}, which supports the use of MACE potentials as force fields through the MLIAP interface, and the data analysis and image generation for the MD simulations were carried out with OVITO~\cite{ZDDP-stukowski2009visualization}.

The amorphous Si$_3$N$_4$ and SiO$_2$ samples were generated through a melt-quench protocol, consisting of an NVT temperature ramp up to 5000 K, a randomization stage at 5000 K, and a two-step (NVT + isotropic NPT) quench back to 300 K and zero pressure, followed by a final structural relaxation; the full protocol is reported in the SI. The structural validation of the resulting amorphous phases against experimental data is presented in Sec.~\ref{sec:validation_apps} and in the SI (Fig.~\textbf{S6}). The simulation setups of the tensile deformation tests and of the CVD-like film growth presented in Sec.~\ref{sec:validation_apps} are also detailed in the SI.

For the sliding simulations, amorphous Si$_3$N$_4$ slabs matching the lateral size of the crystalline diamond slabs were generated by first producing the bulk with the melt-quench procedure and then removing periodic boundary conditions in the z direction. The obtained slabs were annealed at 1500 K for 200 ps and quenched back to 300 K in a 200 ps ramp, in order to heal the surface dangling bonds created by the cut. The interfaces were then assembled by placing the diamond slab on top of the a-Si$_3$N$_4$ slab. Four diamond orientations were considered: C(001), C(110), C(111) and the Pandey-reconstructed C(111) (referred to as C(111)-pd), which is thermodynamically more stable than the unreconstructed surface~\cite{Pandey1981new,Vanderbilt1984total}; each was simulated both clean and with the H+OH passivation (i.e. the termination produced by dissociated water, representative of a humid environment~\cite{leriche2025passivation}). The lateral cells measure approximately 40$\times$40 Å$^2$, for a total of $\sim7000$ atoms; the exact cell sizes and atom numbers are reported in Table~\textbf{S2} of the SI.

The bottom a-Si$_3$N$_4$ slabs (23 Å thick) were partitioned into a lower 5 Å-thick slice kept fixed, a 10 Å region thermostatted at 300 K, and a 7 Å-thick surface region evolving in the NVE ensemble. The diamond slabs ($\sim$13 Å thick) had the topmost atomic layer treated as a rigid body (LAMMPS \texttt{fix rigid}) on which the external normal load and the sliding velocity were applied, a 3 Å-thick NVE region at the surface, and an NVT region at 300 K in between. In this way only regions spatially separated from the shearing interface are thermostatted, while the interfacial atoms evolve according to Newton's equations of motion under the imposed load and sliding.

Each sliding simulation consisted of three stages: (i) a load ramp over 500 ps up to the target normal pressure (0.1, 1, 10 or 23 GPa, the latter corresponding to the hardness of Si$_3$N$_4$; (ii) a sliding velocity ramp from 0 to 10 m/s over 1 ns; and (iii) sliding at constant load and constant velocity $v = 10$ m/s for 7 ns, covering a total sliding distance of 70 nm. We emphasize that these conditions (load, velocity, temperature) fall within the range sampled in the dataset, which included interfaces slid at velocities up to 25 m/s and loads up to 23 GPa. The temperature of the NVE interfacial region was monitored throughout the simulations, remaining stable at 300 K with only random fluctuations of a few K, indicating that at this sliding velocity frictional heating is efficiently dissipated by the thermostats.

The frictional shear stress was computed as the lateral force opposing the sliding motion measured on the driven topmost diamond layer, divided by the cell area, and averaged over the constant-velocity stage of the trajectory. Interfacial bonds were identified geometrically, using species-pair distance cutoffs proportional to the sum of the covalent radii (with a tolerance multiplicative factor), counting both direct C--Si and C--N linkages and oxygen-bridged C--O--Si and C--O--N triplets; linkage densities are normalized by the interfacial area and averaged over the steady-state part of the sliding stage.
Finally, to characterize the localization of shear, we computed planar-averaged velocity profiles $\langle v_x \rangle (z)$ along the direction normal to the interface, averaging over the steady-state sliding. The profiles were fitted with a hyperbolic tangent centered at the slip plane; when the fitted width exceeds a threshold fraction of the profile length, the shear is classified as distributed and the profile is fitted with a linear (Couette-like) function instead.

\section{Results and Discussion}

\subsection{Structural and mechanical properties of the amorphous phases, and film growth}\label{sec:validation_apps}

Before addressing the tribological simulations, we present a set of applications that both further validate the potential on measurable physical properties, and demonstrate its applicability beyond interfacial sliding.

\paragraph{Amorphous structures.} The structures of the melt-quenched amorphous Si$_3$N$_4$ and SiO$_2$ were validated against experimental diffraction data by computing radial and angular distribution functions, reported in the SI (Fig.~\textbf{S6} and Table~\textbf{S1}). For a-Si$_3$N$_4$, the first Si--N peak at 1.74 Å closely matches the experimental value of 1.729 Å for CVD-grown amorphous silicon nitride~\cite{DIAM-Misawa1979}, and the coordination numbers correspond to the expected network of corner-sharing SiN$_4$ tetrahedra connected through threefold-coordinated nitrogen, with a negligible amount of homonuclear bonds. For a-SiO$_2$, the Si--O first peak at 1.63 Å closely matches the experimental distance of 1.62 Å, each oxygen bridges exactly two tetrahedra, and the Si--O--Si angle distribution reproduces the characteristic wide shape peaked around 144$^\circ$~\cite{mozzi1969structure}.

\begin{figure*}[!htb]
\centering
\includegraphics[width=\textwidth]{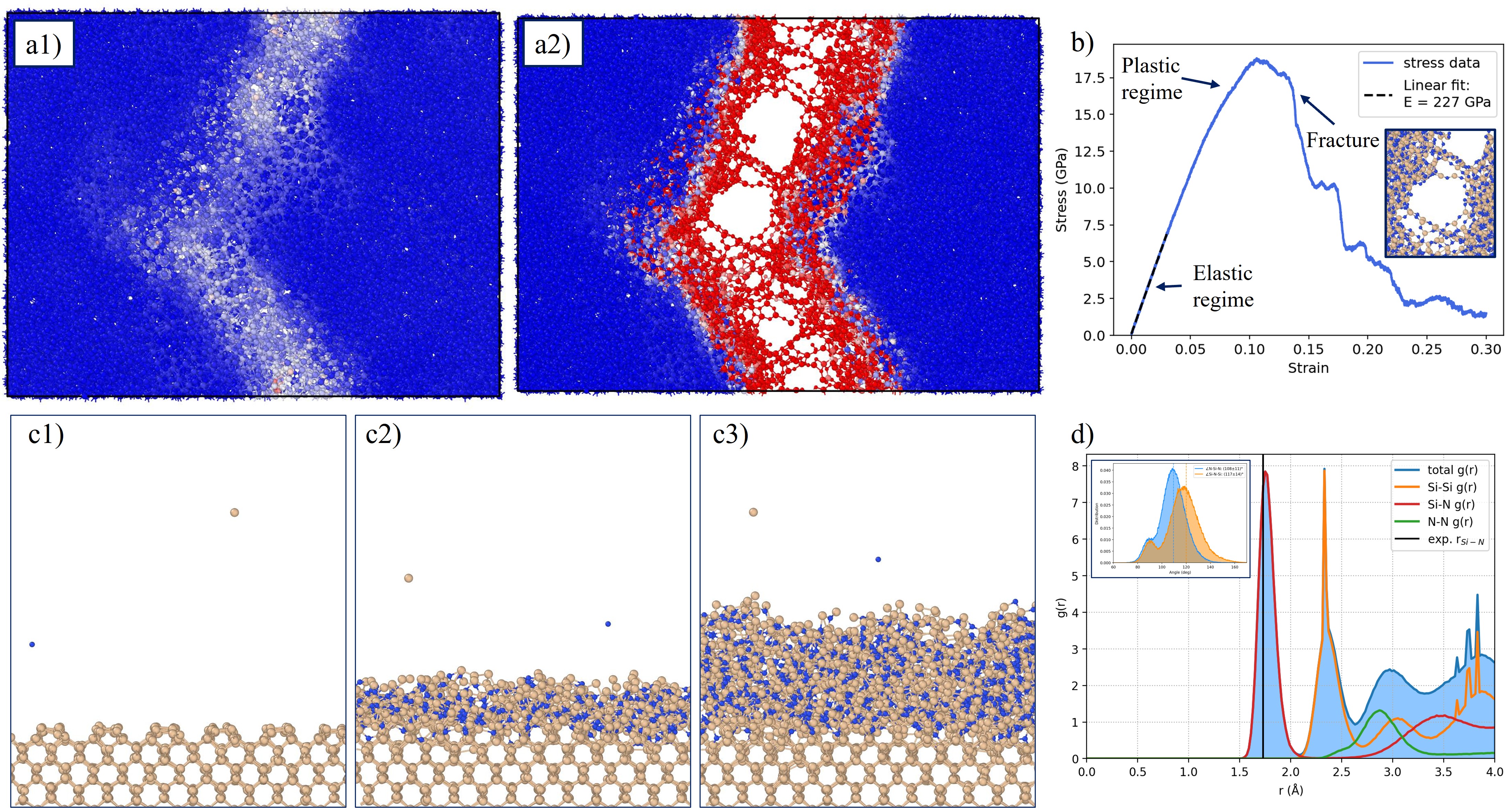}
\caption{Applications of the ML potential to silicon nitride. A tensile deformation test for amorphous Si$_3$N$_4$ is reported in panels a1-a2), color-coded for the von Mises shear strain, showing the initial shear band in a1), ending in fracture at the end (a2), with the associated stress-strain curve in b). CVD-like film growth for (non-stoichiometric) silicon nitride by constant-rate injection of Si and N atoms is reported in panels c1-c3), with the radial pair distribution function (and angular distribution function inset) in d). The RDF peaks are in agreement with experimental values for amorphous silicon nitride, while the sharp peaks are associated to the crystalline Si(100) substrate.}
\label{fig:stress_strain_deposition}
\end{figure*}

\paragraph{Tensile deformation and elastic properties.} As a first mechanical application, we performed tensile deformation tests on melt-quenched amorphous Si$_3$N$_4$ (density 3.0 g/cm$^3$) and SiO$_2$ (density 2.2 g/cm$^3$) samples of $\sim$30000 atoms, strained until fracture at an engineering strain rate of $10^9$ s$^{-1}$ (simulation details in the SI). The stress-strain curve for a-Si$_3$N$_4$ is reported in Fig.~\ref{fig:stress_strain_deposition}b: fitting the initial linear (elastic) regime yields a Young's modulus of 227 GPa, with an associated Poisson ratio of 0.23. This value is in quantitative agreement with the predictions of recent machine-learning potentials developed specifically for a-Si$_3$N$_4$, with moment tensor and GAP models yielding Young's moduli of 220--227 GPa at comparable densities, and with nanoindentation measurements on sputter-deposited films (211--222 GPa)~\cite{nayak2025accurate}. 
Beyond the elastic regime, the sample undergoes plastic deformation with progressive strain softening after a peak stress of $\sim$18.5 GPa at $\sim$11\% strain: the atomic shear strain maps in Fig.~\ref{fig:stress_strain_deposition}a1-a2 reveal the formation of a shear band (a1), which localizes the subsequent deformation, enlarging until the final fracture (a2). The obtained stress-strain curve is in reasonable agreement with previous computational works using classical potentials~\cite{Le2022tensile}. The same test repeated for amorphous silica yields a Young's modulus of 75 GPa (Fig.~\textbf{S7}), in excellent agreement with the experimental value of $\sim$72 GPa for fused silica~\cite{bruckner1970properties}. We note that the imposed strain rate is, as is typical for MD, many orders of magnitude higher than experimental ones, so the plastic branch of the curves should be interpreted qualitatively; the elastic moduli, however, are robust observables that directly probe the curvature of the potential energy surface in the harmonic (elastic) regime around equilibrium.

\paragraph{CVD-like growth of silicon nitride films.} As a second application, we simulated the growth of a silicon nitride film on a Si(100) surface by continuous deposition of atomic Si and N, as a computational analogue of a chemical vapor deposition (CVD) process. Experimentally, silicon nitride films are typically deposited by low-pressure chemical vapor deposition (LPCVD) from dichlorosilane and NH$_3$ at temperatures of 700--800 °C~\cite{gumpher2004characterization, Yang2018characteristic}, through a long sequence of precursor adsorption, dissociation and byproduct desorption events whose overall kinetics is many orders of magnitude slower than the timescales accessible to MD. Following the same rationale as the melt-quench procedure, we therefore accelerated the process by directly depositing the atomic species, bypassing the slow precursor chemistry while retaining the sequential arrival and incorporation of Si and N at the growing amorphous front. A similar strategy is adopted in recent empirical-potential MD studies of plasma-enhanced CVD growth of silicon nitride, in which the cracking of the SiH$_4$ and NH$_3$ precursors is delegated to the (implicit) plasma and pre-dissociated fragments are deposited at comparably accelerated rates~\cite{tian2026growth}. Within this scheme, the relative Si and N fluxes play the role of the precursor flow rates in the experimental process, and the film stoichiometry can be adjusted accordingly: non-stoichiometric (Si-rich) films, which are commonly employed to tune the mechanical (e.g. residual stress)~\cite{zheng2013deposition} and optical~\cite{Lacava2017sirich} properties of nitride coatings, are obtained under Si-rich flux conditions. Besides providing a route to amorphous SiN structures complementary to melt-quench (a supported film with a growth surface, rather than a periodic bulk sample), this simulation probes the potential in the reactive, undercoordinated environments characteristic of a growth front, where the film structure emerges from a long cascade of bond-formation events.

Over 7.6 ns of deposition (setup and injection parameters in the SI), a $\sim$15 nm-thick, continuous and dense film, adherent to the substrate, was grown with a Si-rich composition of N/Si = 1.0 instead of the stoichiometric 1.33 (Fig.~\ref{fig:stress_strain_deposition}c1-c3). Its radial and angular distribution functions (Fig.~\ref{fig:stress_strain_deposition}d), with a Si--N first peak at 1.76 Å and bond angles of N--Si--N $= 108 \pm 11^\circ$ and Si--N--Si $= 117 \pm 14^\circ$, are consistent with the melt-quenched a-Si$_3$N$_4$ and with the experimental values~\cite{DIAM-Misawa1979}; the additional Si--Si peak at 2.33 Å reflects the homonuclear Si--Si bonds expected in a Si-rich film, while the sharp peaks originate from the crystalline Si(100) substrate included in the analysis region.

\subsection{Sliding simulations of diamond--silicon nitride interfaces}

Having established the accuracy and robustness of the potential, we now address the central question of this study: how do diamond surface orientation, reconstruction and passivation determine interfacial bonding and friction under load, over time scales long enough to reach a (nano-)tribological steady state? To this end, we simulated the full matrix of four diamond orientations (C(001), C(110), C(111), C(111)-pd), two terminations (clean and H+OH-passivated) and four normal loads (0.1, 1, 10, 23 GPa), each sliding for 70 nm, as described in the Methods. The clean surfaces are representative of a dry, inert environment or vacuum, in which passivating species are unavailable or were previously removed by sliding and not replaced, while the H+OH termination represents the partial passivation produced by dissociated water in humid ambient conditions. The highest load, 23 GPa, equals the hardness of Si$_3$N$_4$ and thus corresponds to the local pressure experienced by plowing diamond asperities.

\begin{figure*}[!htb]
\centering
\includegraphics[width=\textwidth]{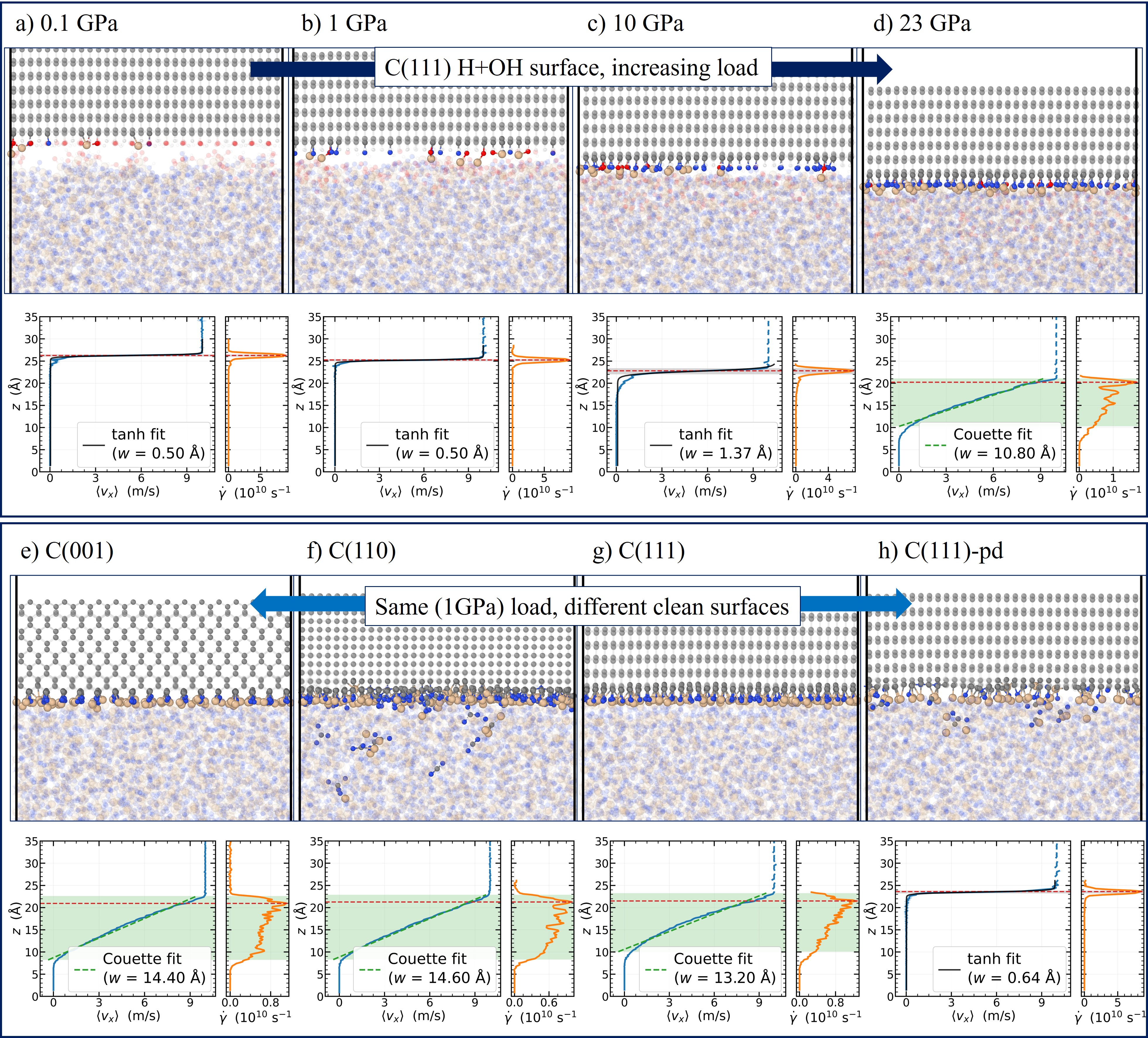}
\caption{MD simulations of diamond-silicon nitride interfaces sliding under an applied load. The top panel (a-d) contains the C(111)--H+OH passivated surface for increasing normal loads, highlighting interfacial bonds and the (planar-averaged) velocity distribution in the sliding direction, along the vertical profile. Up to very high loads, surface passivation prevents chemical bonding across the interface, maintaining the shear plane localized at the diamond-silicon nitride interface. At very high load (23 GPa, equal to Si$_3$N$_4$ hardness) strong interfacial bonding moves the shear plane away from the interface, realizing a Couette-like flow in the Si$_3$N$_4$ region near the interface. The bottom panels (e-h) contain the simulations for the various clean (non-passivated) surfaces at 1 GPa load. For C(001), C(110) and C(111) full interfacial chemical bonding is observed, as for the passivated C(111) at 23 GPa, while the Pandey-reconstructed C(111) forms fewer bonds and maintains the shear plane localized at the interface.}
\label{fig:interfaces_grid}
\end{figure*}

\begin{figure*}[!htb]
\centering
\includegraphics[width=\textwidth]{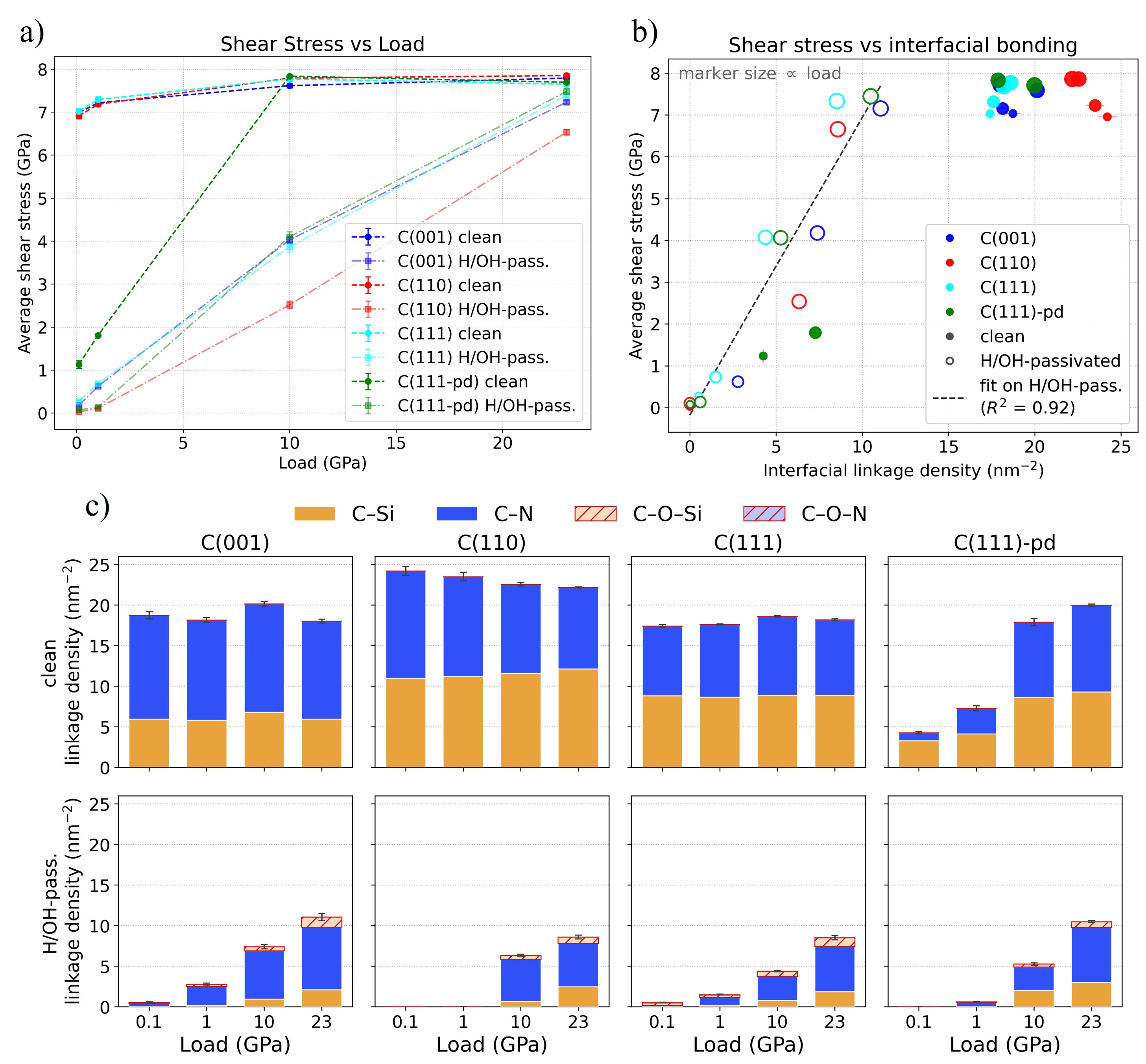}
\caption{Trends of the shear stress as a function of load a), shear stress as a function of interfacial bonds (b), and bond density (c) for the different combinations of diamond surface orientations, passivations, and applied load. For the non-passivated surfaces, full bonding occurs in all conditions (except for the C(111)-Pandey reconstructed at low loads), yielding comparable shear stresses, with weak load dependency, dominated by the shear plasticity of Si$_3$N$_4$. For the passivated surfaces, a clear relation is established between shear stress, interfacial bond formation, and load.}
\label{fig:interfacial_analysis}
\end{figure*}

\paragraph{Effect of passivation and load.} Figure~\ref{fig:interfaces_grid}a-d illustrates the behavior of the H+OH-passivated C(111) surface with increasing load, showing MD snapshots with the interfacial bonds highlighted, together with the planar-averaged velocity profiles $\langle v_x \rangle(z)$ and their derivatives. At the two lowest loads (0.1 and 1 GPa), the passivation layer keeps the surfaces chemically separated: essentially no covalent linkages form across the interface over the entire 70 nm of sliding, and the velocity profile is a sharp step located at the interface, with a fitted slip-zone width smaller than 1 Å, corresponding to the two bodies sliding rigidly past each other along a well-defined, atomically sharp slip plane. At 10 GPa, a finite density of interfacial linkages develops ($\sim$2.7 nm$^{-2}$, see below), formed by the mechanically activated reaction of surface groups; the shear plane, however, remains localized at the interface (width 1.4 Å), as the bonds are continuously broken by sliding. Only at the highest load of 23 GPa interfacial bonding becomes extensive enough ($\sim$8 nm$^{-2}$) that the interface becomes more resistant to shear than the amorphous nitride itself: consequently, the shear plane moves away from the interface and the deformation is redistributed within the a-Si$_3$N$_4$ film, whose velocity profile becomes almost linear over a width of 11 Å, corresponding to a Couette-like flow of the amorphous solid.
We also note that a fraction of the linkages formed by the passivated surfaces are oxygen-bridged (C--O--Si and C--O--N), i.e. mediated by the hydroxyl groups after the transfer of their hydrogen to the nitride, in agreement with the Si--O--C bonding pathway observed in the AIMD simulations of our previous work~\cite{leriche2025passivation}.
A closely analogous, load-activated mechanochemistry has recently been reported for H/OH-passivated amorphous carbon and silica sliding on epitaxial graphene: below thresholds of ${\sim}10{-}12.5$ GPa the passivation keeps the interface chemically inert, while above them the H and OH groups transfer to the graphene and interfacial bonds form, in part through oxygen-bridged linkages~\cite{zarshenas2025transient}.

\paragraph{Effect of surface orientation and reconstruction.} The bottom panels of Fig.~\ref{fig:interfaces_grid} (e-h) compare the four clean surfaces at a normal load equal to 1 GPa. Here the picture is reversed: for C(001), C(110) and C(111), the absence of passivation leads to essentially complete chemical bonding across the interface already at lower load, with linkage densities of 18--24 nm$^{-2}$; correspondingly, the interface is stronger than the nitride and the shear is distributed, with Couette-like velocity profiles spanning most of the mobile thickness of the a-Si$_3$N$_4$ film. The notable exception is the Pandey-reconstructed C(111), whose $\pi$-bonded chain reconstruction~\cite{Pandey1981new,Vanderbilt1984total} makes the surface more inert: at 1 GPa it forms a much lower density of interfacial bonds ($\sim$7 nm$^{-2}$), and the shear plane remains localized at the interface.
Notably, for the clean C(110) a few carbon atoms are detached from the surface during sliding and incorporated into the nitride, in line with the known higher susceptibility of the C(110) surface to tribochemical wear observed in previous studies~\cite{DIAM-peguiron2016activation, DIAM-Ta2023}. This behaviour is also consistent with the surface energies of the different facets (5.9 and 4.1 J/m$^2$ for the C(110) and C(111)-pd, respectively)~\cite{stekolnikov2002absolute}.

\paragraph{Shear stress trends.} The average shear stresses for all 32 simulations are collected in Fig.~\ref{fig:interfacial_analysis}a as a function of load. The clean C(001), C(110) and C(111) interfaces exhibit a strikingly weak load dependence: the shear stress is already $\sim$7 GPa at a nominal load of only 0.1 GPa, and saturates at 7.6--7.9 GPa at the highest loads. This indicates a sliding regime where the resistance to shear of the diamond-silicon nitride interface, due to full chemical bonding, is stronger than the intrinsic shear resistance of amorphous nitride itself, so that the shear stress is determined by plastic flow of the amorphous region near the interface. As strain rate is fixed for all the cases, the small load dependence of shear stress can then be interpreted in analogy to the weak load dependence of viscosity for laminar flow in Newtonian fluids. 
The same mechanics has been observed in DFTB sliding simulations of SiC-silica and SiC-DLC interfaces: once interfacial bonds form, the shear plane relocates into the softer of the two bodies, whose yield stress decreases under pressure, and the resulting high-pressure shear-stress plateau is only marginally sensitive to the sliding velocity~\cite{zarshenas2025transient}.

The clean C(111)-pd interface again behaves differently: at 0.1 and 1 GPa its shear stress is only 1.1--1.8 GPa, reflecting the reduced bonding due to the reconstruction, but at 10 GPa and above the protective effect of the $\pi$-bonded chains is overwhelmed, the reconstruction is lost and bonding becomes complete (Fig.~\ref{fig:interfacial_analysis}c), which results in the shear stress joining the plateau of the other clean surfaces. Notably, the behaviour at lower loads is consistent with AIMD simulations of diamond sliding against silica at 1 GPa~\cite{giaremis2025tribological}, which showed only non-bonding interactions at the Pandey-reconstructed C(111).
The passivated interfaces, in contrast, show a strong and monotonic load dependence, in line with the Amontons--Coulomb for friction: the shear stress grows from nearly zero at 0.1 GPa to 2.5--4.1 GPa at 10 GPa, and finally approaches the clean-surface plateau (6.5--7.5 GPa) at 23 GPa, where passivation is no longer able to prevent extensive interfacial bonding. Among the passivated surfaces, C(110) is consistently the best protected, forming no measurable linkages up to 1 GPa and retaining the lowest shear stress at all loads.

\paragraph{Interfacial bond composition.} The composition of the interfacial linkages, distinguishing direct C--Si and C--N bonds from the oxygen-bridged C--O--Si and C--O--N linkages, is reported in Fig.~\ref{fig:interfacial_analysis}c for all cases. For the clean surfaces (top row), the total linkage density is high (18--24 nm$^{-2}$) and only weakly dependent on load, confirming that bonding is saturated already at 0.1 GPa; C(110) systematically forms the largest number of bonds, consistent with the high density of reactive dangling bonds exposed by its unreconstructed zigzag chains, and C--N linkages outnumber C--Si ones on all orientations (most markedly on C(001), where they are roughly twice as abundant). The C(111)-pd surface is again the outlier, with its linkage density growing from $\sim$4.5 nm$^{-2}$ at 0.1 GPa to $\sim$20 nm$^{-2}$ at 23 GPa, mirroring the load-induced breakdown of the reconstruction discussed above. For the passivated surfaces (bottom row), the linkage density grows monotonically with load for all orientations, from zero (or nearly zero) at 0.1 GPa to 8--11 nm$^{-2}$ at 23 GPa -- still well below the clean-surface values, showing that even at the plowing pressure the passivation removes roughly half of the interfacial bonds, at least within the sliding distances covered by the simulations. 

\paragraph{Relation between friction and interfacial bonding.} The central result of this analysis is obtained by plotting the shear stress directly against the interfacial linkage density, as done in Fig.~\ref{fig:interfacial_analysis}b for all 32 simulations. Two regimes clearly emerge. In the first, bond-controlled regime, populated by the passivated interfaces (and by the clean C(111)-pd at low load), the shear stress is proportional to the linkage density: a linear fit restricted to the passivated cases yields $R^2 = 0.92$, with a slope of $\sim$0.7 GPa/nm$^{-2}$.

Such a linear trend suggests a simple mechanistic picture for friction in this regime, in which resistance to sliding motion is due to the continuous interfacial bond breaking and formation, consistent with the theoretical model proposed by Filippov et al.~\cite{filippov2004friction}. Additionally, since the interfacial bonding events are the precursors of atom-by-atom carbon removal, this regime establishes a direct link with the elementary processes responsible for diamond wear.
This picture is corroborated by recent reactive MD simulations of the atomic-scale wear of DLC-DLC interfaces, in which the number of wear events was found to grow linearly with the number of interfacial bonds \cite{huang2025reactive}, and wear was proposed to proceed in two steps: interfacial bond formation, activated mainly by the normal stress, followed by removal of the bonded atoms, activated by the shear stress~\cite{DIAM-wang2020}.
A direct correlation between interfacial bond count and shear stress has also been reported for passivated a-C sliding on epitaxial graphene, with load-activated bond densities (${\sim}2{-}8$ nm$^{-2}$) closely matching the range found here for the passivated interfaces~\cite{zarshenas2025transient}.
In the second, saturated regime, populated by the clean surfaces, the data points cluster at high linkage densities (17--25 nm$^{-2}$) and a nearly constant shear stress of 7--8 GPa, independent of the bond count: once the interface is stronger than the a-Si$_3$N$_4$, additional bonds no longer increase friction, because the weakest shear plane has moved from the interface into the amorphous solid, and the stress is capped by its plastic flow.

\paragraph{Implications for diamond wear.} These results carry general implications for the wear of diamond-ceramic contacts. First, interfacial covalent bonding -- the necessary first step of atom-by-atom diamond wear -- is essentially absent on H/OH-passivated surfaces up to loads of the order of 1 GPa, i.e. over most of the pressure range of a multi-asperity contact, and passivation loses its protective role only approaching the plowing pressure set by the hardness of the softer body: this provides a quantitative, load-resolved basis for the environmental sensitivity of diamond wear rates observed experimentally~\cite{DIAM-konicek2012influence,DIAM-ta2023effects,leriche2025passivation}.
Moreover, the reactivity hierarchy of the diamond facets is now quantitatively established, from the highly reactive C(110), which forms the largest number of interfacial bonds and releases carbon atoms into the counter-body, to the Pandey-reconstructed C(111), intrinsically protected up to several GPa even without passivating species: since polycrystalline coatings expose a distribution of facets, this implies crystallite-to-crystallite variability of the wear rate.

\section{Conclusions}
In this work, we developed a general-purpose machine learning potential for the C/H/N/O/Si chemical space, describing diamond and DLC surfaces, silicon nitride, silica and silicon in crystalline and amorphous form, environmental molecules, and their interfaces under load and shear. Two MACE models were trained on more than 30,000 DFT configurations of 253 systems: a fine-tuned model (MACE-ft) for the highest accuracy, and a lighter model (MACE-small) with nearly the same accuracy at a much lower cost. Both reproduce the DFT dissociation barriers of H$_2$, H$_2$O and O$_2$ and carbon-chain peeling energetics on diamond surfaces, the phonon dispersions of the crystalline phases and the experimental structure and elastic moduli of amorphous Si$_3$N$_4$ and SiO$_2$. Tensile-to-fracture tests and the CVD-like growth of a Si-rich silicon nitride film show its applicability beyond interfacial sliding.

Sliding simulations of diamond--Si$_3$N$_4$ interfaces, covering four surface orientations, clean and H/OH-passivated terminations and loads from 0.1 to 23 GPa, reveal two friction regimes distinguished by the density of interfacial bonds. On passivated surfaces, and on the Pandey-reconstructed C(111) up to about 1 GPa, few bonds form, sliding is localized at an atomically sharp interfacial plane, and the shear stress grows linearly with the bond density. On the clean C(001), C(110) and C(111) surfaces, bonding is complete at all loads, shear moves into the nitride, which flows in a Couette-like manner, and the shear stress saturates to the limit set by the plastic flow of amorphous Si$_3$N$_4$. The passivated surfaces approach this plateau only at 23 GPa, with about half the bond density of the clean ones, and the Pandey reconstruction is lost from 10 GPa.

These results show how passivation, surface orientation and load together determine interfacial bonding and friction. Since interfacial bonding is the first step of atom-by-atom wear, they also provide a load-resolved basis for the environmental sensitivity of diamond wear: H/OH passivation prevents bonding over most of the pressure range of a multi-asperity contact, and loses its protective role only near the plowing pressure. More broadly, the potential enables reactive simulations beyond the length and time scales of first-principles methods, for example of wear at plowing diamond nanoasperities, diamond--silica interfaces relevant to chemical-mechanical polishing, molecules intercalated in the contact, and the growth and mechanical response of carbon and silicon nitride films.

\section*{acknowledgement}
These results are part of the TRIBOSYNC project, which received funding from the European Research Council (ERC) under the European Union’s HORIZON.1.1 Proof of Concept program. (Grant Agreement No. 101292653). \\
We acknowledge EuroHPC-JU for awarding proposal ID: EHPC-REG-2025R01-165 with access to HPC resources on JUPITER Booster at Jülich(Germany) \\
We also acknowledge the CINECA award under the ISCRA initiative, for the availability of high- performance computing resources and support.
We thank Dr.~Bart Weber (Advanced Research Center for Nanolithography, ARCNL) for insightful discussions and for the ongoing collaboration on diamond wear experiments that motivated this work.

\bibliography{bibliography}

\end{document}